# Lifetime Sample Tracking (LiST): A Data Platform for Materials Science

Anthony Richardella[1,6], Isaiah A Moses[1], Konrad Hilse[1], Frank Santaguida[1], Kevin Dressler[1], Ric Wilburn[2], Saiyyam Kochar[3], Wesley F. Reinhart[1,4,7], Adri C. T. van Duin[1,4,5], Nitin Samarth[1,3,4,6], Vincent H. Crespi[1,6], Joan M. Redwing[1,4]

[1]*2D Crystal Consortium Material Innovation Platform (2DCC-MIP) Materials Research Institute, The Pennsylvania State University, University Park, Pennsylvania 16802, United States*

[2]*The Center for 3D Ferroelectric Microelectronics Manufacturing (3DFeM[2]), The Pennsylvania State University, University Park, Pennsylvania 16802, United States*

[3]*Q-NEXT, National Quantum Information Science Research Center, Argonne National Laboratory, Lemont, IL, 60439, United States*

[4]*Department of Materials Science and Engineering, The Pennsylvania State University, University Park, Pennsylvania 16802, United States*

[5]*Department of Mechanical Engineering, The Pennsylvania State University, University Park, Pennsylvania 16802, United States*

[6]*Department of Physics, The Pennsylvania State University, University Park, PA, USA*

[7]*Institute for Computational and Data Sciences, The Pennsylvania State University, University Park, PA 16802, United States*

The 2D Crystal Consortium Materials Innovation Platform (2DCC-MIP) is an NSF supported national user facility focused on advancing the synthesis of 2D materials, monolayers, surfaces, and interfaces. The need for the facility to organize and share data with users led to the development of an internal data management and analysis engine, the Lifetime Sample Tracking platform (LiST). This infrastructure allows the automated capture, curation, analysis and dissemination of data ranging from experimental materials synthesis parameters and characterization, to theoretical first-principles and ReaxFF molecular dynamics modeling[1]. The system currently hosts synthesis and property data (accessible via a REST API) on approximately twenty thousand samples produced by the 2DCC grown using a variety of techniques from bulk crystal growth to metal-organic chemical vapor deposition (MOCVD) and molecular beam epitaxy (MBE), among others. Data used in publications can easily be grouped by the system into data packages that are given digital object identifiers (DOIs) for inclusion with each publication. The LiST platform is now being used by groups outside of the 2DCC as a solution for data curation in materials science. Data management tools such as LiST support the materials development process by allowing a closed loop iteration between synthesis, characterization, theory, and targeted materials design. This also enables machine learning (ML) research, artificial intelligence (AI) analysis, and the potential for autonomous synthesis in the future.

# Motivation

The Materials Genome Initiative (MGI) was established with the idea of taking the complexity that exists in materials, figuring out the key parameters that give desirable properties in a class of materials and using those to go from a research and development mode of trial and error to one of materials-by-design. This envisioned taking the vast quantity of research results that are produced each year and making the data behind them findable, accessible, interoperable and reusable (FAIR)[2]. Since then, repositories of computational materials science such as AFLOW[3], Materials Cloud[4] and Materials Project[5] have become highly successful. In the experimental space, in narrowly defined domains, efforts like the ICSD[6] database of inorganic crystal structures have demonstrated the value and utility of repositories of well curated experimental data.

The totality of experimental materials science, however, entails complexity that does not lend itself to a unified standard organizational hierarchy. The vast number of possible material compounds and physical forms they can take are mirrored by numerous techniques used to synthesize and study them. Further, while theoretical materials science can typically assume well-defined crystal structures, real materials are often a product as much of their defects as their crystal structure and their properties depend intimately on the details of how they were created. The synthesis steps and parameters need to be recorded meticulously and consistently, but practically this is often difficult. Parameters are highly specific to particular techniques. Instrumentation is typically a commodity with many possible manufacturers, commonly using proprietary software and incompatible data formats. Human record keeping is often inconsistent among researchers and error-prone due to manual entry as a common practice. Further, experimental material scientists typically operate in small, short-term collaborations among labs with publications involving a handful of authors. The record keeping of these groups may not consider the broader long-term needs of the community, in contrast to fields with larger collaborations where such standards naturally arise. Finally, publications tend to focus on the best samples produced and failed or sub-optimal results are discarded with little mention in the literature. This necessitates tools and standards in data management to make the data useful according to the FAIR Data Principles for future re-use.

The rise of ML and AI has highlighted these deficiencies, such as the value in preserving negative results. Good and bad results are needed to train the algorithms to predict better synthesis parameters. The experimentalist's task in processing and curating just good data can already feel daunting enough. Tools like laboratory information management systems (LIMS) can address this by automatically capturing data and metadata in a way that can be understood and reused. The LiST platform grew out of the need to manage and share data with users of the 2DCC facility and, as shown in Fig. 1(a), it is a core part of the 2DCC's mission to serve the materials research community. LiST was an organic product of the needs the 2DCC encountered within the realm of the synthesis of 2D materials for a data management system that was more complex than the requirements of an individual research group, but more focused than the federated systems employed at large national labs[7–9]. An observation in its development was that archived data is not useful until it is used. That is, the more data captured in a system like LiST solves the needs of researchers, the more it is used by them, and the more gaps in the collection process become clear and urgent enough for researchers to want to address. This is how data culture begins to change. The platform is grounded in the principles of FAIR data but also shaped by the pragmatic requirement of having to meet the needs of multiple research groups within the 2DCC, each with their own highly specialized instruments and tools, spanning growth, characterization and theory and simulation. We believe this gives the platform the potential for broad utility to the community beyond a particular synthesis technique or set of materials.

Work is currently underway to expand its use in partnerships with the Penn State 3DFeM[2] Engineering Frontier Research Center and Argonne's Q-NEXT center. These partnerships are adding the needs of the centers beyond the 2DCC into the LiST architecture. New features are continuously being developed, such as expanding the lifecycle of materials to include substrate inventories and substrate activities (e.g., polishing, cleaving, growth) so they can be tracked just as produced samples, recognizing that substrate origin, vendor details and processing of the substrate prior to growth are as important as capturing synthesis and characterization.

The system contains growth and characterization data on 15,000 sample growths to date that were synthesized by metal organic chemical vapor deposition (MOCVD), molecular beam epitaxy (MBE) and a variety of bulk growth or other techniques. From these, pieces have been split or cleaved to produce approximately 20,000 unique pieces of material that are tracked by the system, many of which were shipped to users of the 2DCC. Associated *in-situ* and *ex-situ* characterization data, such as AFM, ARPES, PL, Raman, RHEED, transport, TEM, and XRD are captured for each sample. Core features of LiST are the automated collection of growth recipes and parameters from synthesis instruments, collection of characterization data, standardized naming conventions for easy search, integration of simulation and machine learning tools, sharing of data with external collaborators, use of DOIs for physical instruments to include in methods sections of publications, and creation of Data Packages that auto-generate DOIs for preserving data used in publications.

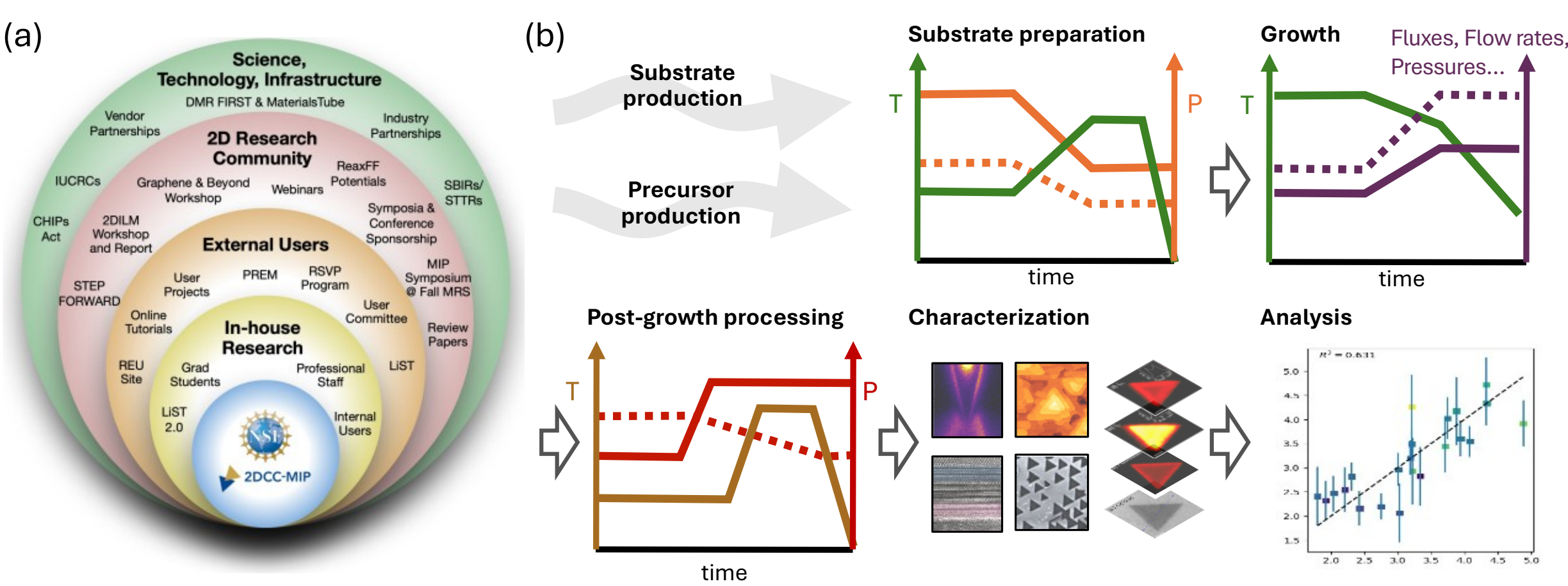

*Fig. 1 (a) Overview of the mission of the 2DCC-MIP in its first 10 years of operation. Development of tools like LiST are one part of its broader mission to serve the materials research community. (b) To obtain transformative data-driven insights into materials growth, the data and metadata that represents a material must include not only the measurements characterizing its current state, but also the full history of how it was produced. These data must be collected at scale, to effectively cover the large parameter space of growth protocols and growth chamber designs that produce materials.*

# The LiST Database

As illustrated in Fig. 1(b), there are details at every step of the research process that need to be properly recorded to understand how a sample resulted with the properties it has. LiST aims to capture the full history of how each sample was synthesized, the instrumentation used, the characterization performed on it and the final publication outputs that result from it. The LiST database contains over 1000 unique materials and heterostructures at the time of writing. These were grown using various methods including bulk growth methods, such as Bridgman, flux growth, floating zone and chemical vapor transport (CVT); thin film methods, such as MBE, hybrid physical-chemical vapor deposition (HPCVD), and MOCVD; and confinement heteroepitaxy (CHet) using Epigraphene (graphene on SiC). It should be noted that many of the individual samples captured are non-ideal or "bad" growths, which are needed to train models to distinguish good and bad regions of parameter space when predicting optimal conditions. The LiST operating model is to record all samples and characterization data regardless of the perceived value of the data.

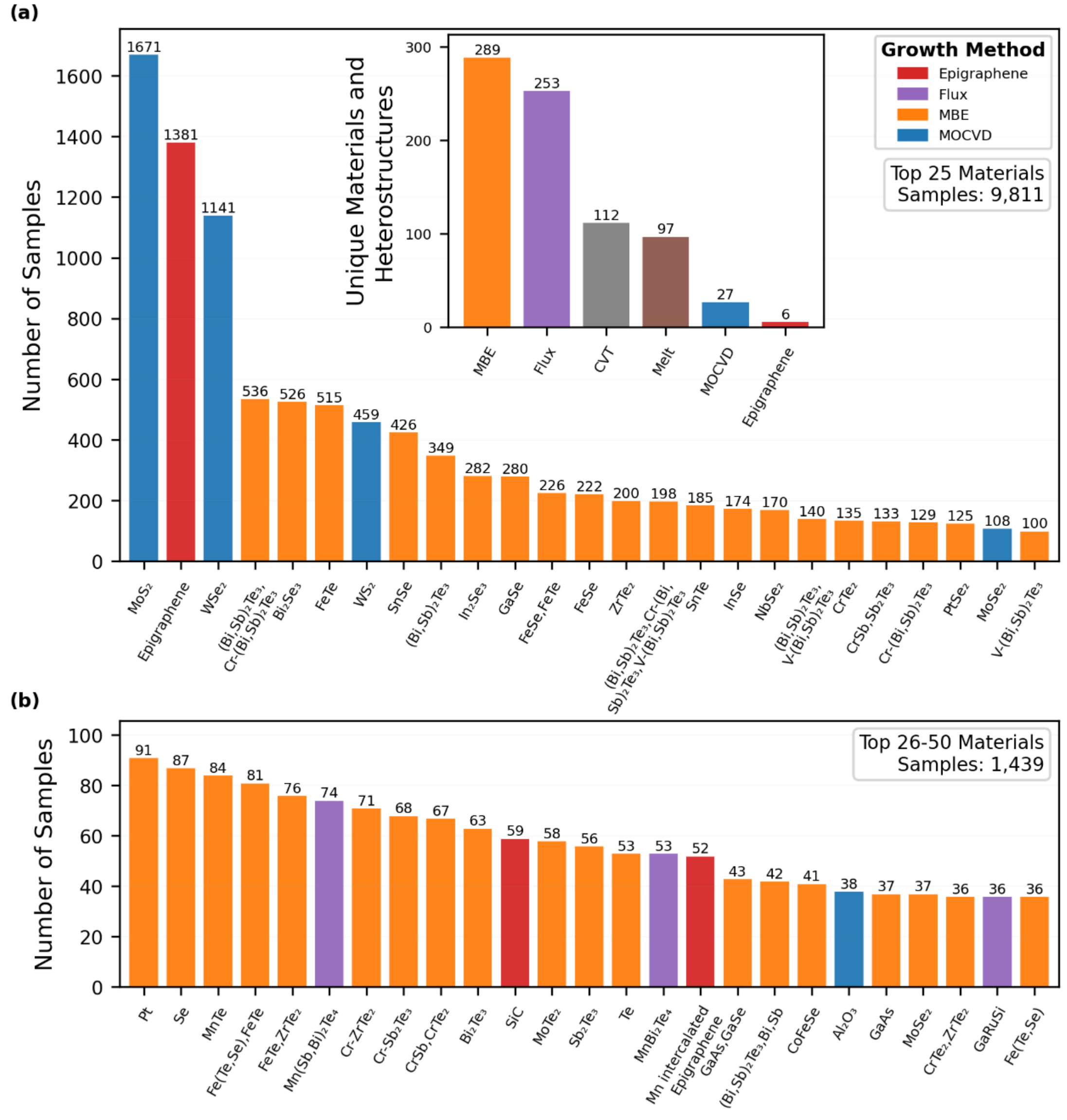

*Fig. 2 Materials with the highest number of samples in LiST, categorized by their growth methods. **(a)** displays the top 25 materials by number of samples, while **(b)** shows the next 26-50$^{th}$ top materials. The inset in (a) illustrates the total number of unique materials and heterostructures in LiST synthesized using different growth methods.*

The top 50 materials in terms of numbers of samples are highlighted in Fig. 2. MBE synthesis produced the largest number of samples, while bulk growth methods have been applied to the greatest variety of materials. Transition metal dichalcogenides (TMDs) are the most studied material type in the LiST database, with MOCVD growth of $MoS_2$ having the largest number of samples, with $WSe_2$ and $WS_2$ close behind. Other materials span topological insulators (TI) and magnetic TIs (e.g. $Bi_2Se_3$, Cr doped $(Bi, Sb)_2Te_3$, and $MnBi_2Te_4$), iron-based superconductors (e.g., FeSe and FeTe heterostructures), 2D magnets (e.g., $CrTe_2$ and $Cr_2Ge_2Te_6$), and altermagnets (MnTe and CrSb). Examples of the types of material classes can be seen in Fig. 3. The full list of public data can be viewed at https://data.2dccmip.org/list/public/sampleMetaData.

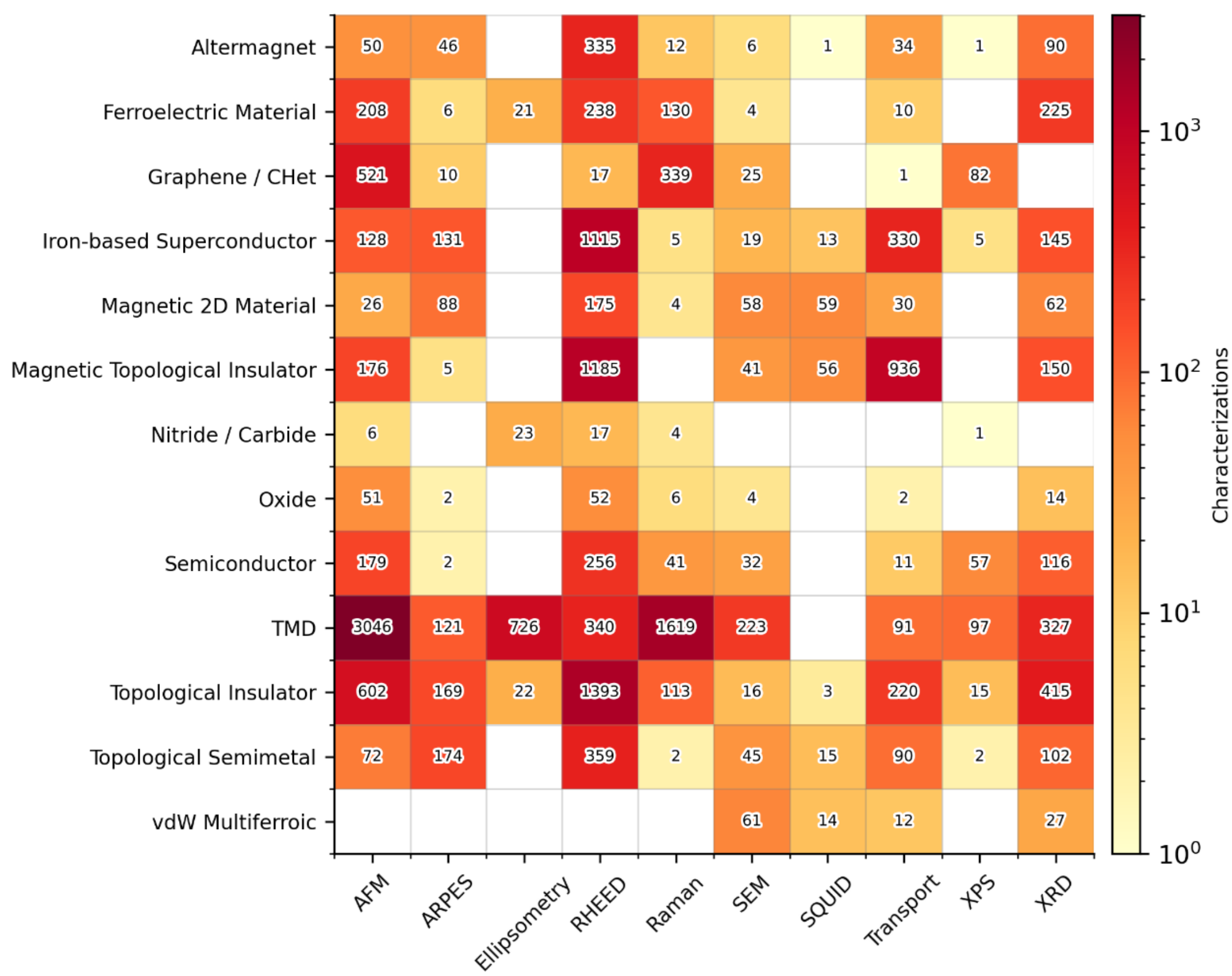

*Fig. 3. Heatmap displaying statistics of studies by characterization technique of samples in various material categories, with darker colors indicating the highest data density and cases with no data in white. Each characterization activity typically produces multiple individual measurements, so the total number of data files is larger than the number of studies shown here.*

Various characterization techniques were employed to study the properties and qualities of the synthesized materials. The techniques applied to the most samples include reflection high-energy electron diffraction (RHEED) with ~5,700 (mostly MBE) samples, atomic force microscopy (AFM) with ~5,200 samples, Raman spectroscopy and photoluminescence with

~2,350 samples, X-ray diffraction (XRD) and reflectivity with ~1,950 samples, and electrical transport measurements with ~1,900 samples. Other techniques have been used on hundreds of samples, including scanning electron microscopy (SEM), ellipsometry, angle-resolved photoemission spectroscopy (ARPES), X-ray photoelectron spectroscopy (XPS), and UV-visible (UV-Vis) spectroscopy. The relative number of characterization data for the most populated materials is shown in Fig. 4.

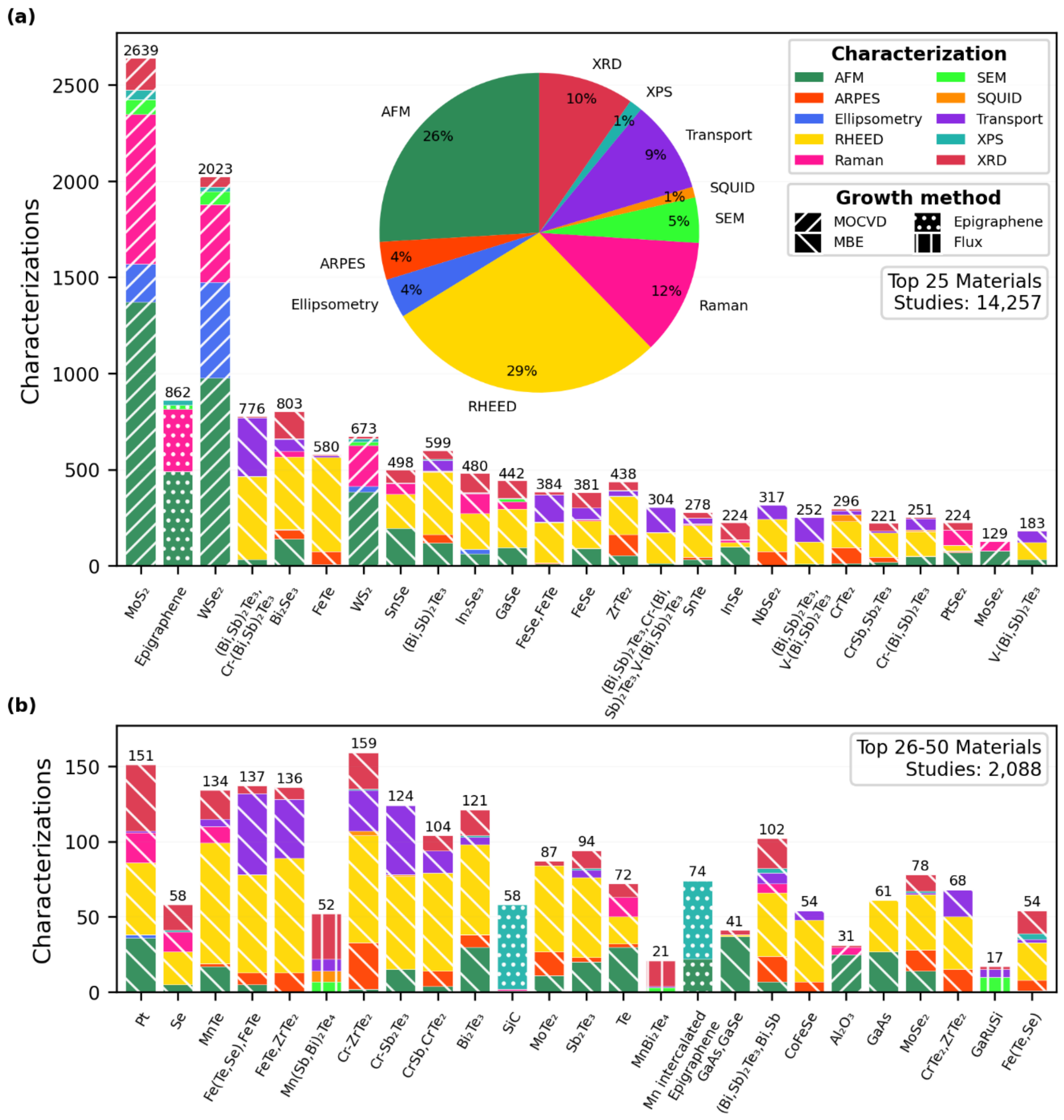

*Fig. 4. Characterizations of materials with the highest number of samples.* ***(a)*** *illustrates the characterizations of the top 25 materials, while* ***(b)*** *focuses on the next 26-50$^{th}$ top materials. Pie chart inset in (a) represents the instances of the most common characterization techniques as a percentage of all characterizations done on all materials.*

# Capabilities and Architecture

The history of each of these samples spans from how their substrates and precursors were prepared, through growth and processing conductions, to the analysis of characterization and results included in publications. These details may be very different for different growth techniques or even different synthesis chambers of the same technique. This necessitates a system design that is modular and extensible, as future requirements cannot be predicted *a priori*. The LiST system architecture includes three modular components, 1) the LiST website, database and API, 2) data collection from instrumentation to cloud storage and 3) data processing, theory and simulation tools.

**LiST Application and API**

Fig. 5(a) shows samples associated with a user's project in the LiST interface, including sample growth parameters, a table of timestamped growth log events, and links to all characterization activities for each sample (with some representative data illustrated at right). Additionally, the system allows initiating and tracking events such as shipping a sample to an external user, thereby providing a sample providence capability. Users receiving samples from the 2DCC can log into the LiST website to view and download data for their projects and samples and are encouraged to upload their own data acquired at their home institutions if they choose to. Data scientists can obtain direct access to LiST data via API, described below, through a Data Request.

The LiST server is described schematically in Fig. 5(b) with various APIs and services implemented in a .NET framework. The front-end is a React-based web application implemented using the Next.js framework providing fast data binding between the system front-end and the back-end database. The background service is responsible for ingesting new files and metadata into the system as they are saved in a Microsoft SharePoint cloud storage site using notifications from the Microsoft Graph API, which are then indexed in the SQL database.

LiST is available to both Penn State users and external collaborators. Users authenticate using either Penn State's Microsoft Entra ID via the Microsoft Authentication Library (MSAL) or using the Shibboleth InCommon Federation. Both Entra ID and Shibboleth are solely used to verify user identity; LIST then maps the authenticated user's identifier to an internal user record if one exists and issues a signed secure access token (JWT) reflecting the user's permissions in LiST. LiST also allows anonymous read-only access to published data without authentication, and authenticated users that are unknown to LiST are also only given read-only access to public data.

External users may request access by generating an API key through a self-service form, providing their name, email, and institutional affiliation. These keys also allow read-only access to all published data packages via the API. Authenticated collaborators with LiST accounts can also generate API keys that are linked to their account permissions and provide access to both public data and any private data they have been authorized to view. Users with a research interest in unpublished datasets may submit a data request describing the intended use. If approved, a dedicated project is created, and the relevant samples are linked accordingly. The requester can then generate an API key with access limited to the approved data.

The LiST API provides programmatic access to metadata and datasets managed within the system using a RESTful API. It is designed to support data reuse, interoperability, and integration with external analysis tools and research workflows, enabling machine learning analyses of the data in LiST[10–14]. The API supports metadata-based queries across a range of fields, such as sample composition, synthesis parameters, and project identifiers. Authenticated users can query their sample records, retrieve associated data files, and explore data packages in a structured, machine-readable format. For publicly available published datasets, citation information and persistent identifiers (DOIs) are included to support proper attribution and reproducibility.

**Automated Data Collection**

The second component of the LiST system (Fig. 5(c)) automates the collection of data and metadata so they can be ingested by the LiST server. This is designed with the observation that software for our instrumentation was overwhelmingly Windows OS based and that the experimental researchers using them were most familiar with that environment. Additionally, it was beneficial to keep instruments isolated from public networks to avoid issues with updates, vulnerable software, or other security issues. Finally, implementing automation drastically reduced human error and inconsistent data. Therefore, labs were setup with a Windows "Gatekeeper" computer where all instruments in that lab transfer data, logs and other information over an isolated local network. The Gatekeeper is responsible for uploading data to cloud storage, where LiST can ingest it. This preserves the independence of each lab, provides a single pathway to export data from a lab, and reduces the number of computers that need to be configured to communicate with LiST. The Gatekeeper also provides edge computing resources to each lab for latency sensitive AI/ML analyses.

Where possible, metadata capturing software was integrated into the control systems of synthesis computers so that initiating a sample growth prompts the user for details of what material they intend to grow, for whom, the substrate used, etc. and then automatically assigns a standardized unique sample ID that is used for all subsequent activities with that sample. The software saves a reliable sample start time so that time-stamped instrument logs can provide a record of actual growth parameters. In cases where this integration is not possible, records and sample IDs can be generated on the LiST website. When necessary, Windows symbolic links or hard links were used to move log locations from instrument computers to the Gatekeeper so they are captured. Characterization data, in contrast, are often acquired in shared facilities over which the 2DCC has limited control. These instruments were given standardized names and folders on SharePoint where users were trained to place their data in folders labeled with the sample ID. Compliance is mandated by policy and motivated by providing automated data analysis and a standardized location in LiST where sample data can be found, rather than the user having to manage and organize it themselves. Incomplete or mislabeled data is flagged by the system for review.

**Data Processing and Simulation**

The data processing, theory, and simulation module is currently the least integrated of the three modules and one of the most active areas of development. Automated exports of binary data files to text files, and basic analysis routines are run automatically. The 2DCC is also integrating state of the art theoretical modeling tools into the LiST platform. Whereas excellent national resources exist for equilibrium properties of materials, materials growth kinetics are currently not supported by similar cyberinfrastructure resources. Recognizing this, ReaxFF[1] reactive force field parameter sets have been made available to the community through the LiST interface at https://data.2dccmip.org/reaxff. This provides a centralized repository of potentials that have been developed for a range of elements, allowing wider access to kinetic growth simulations. Results from other techniques, such as first-principles DFT simulations, are archived in LiST results in much the same way that experimental sample data is. Work is ongoing to integrate Globus pipelines to pull data from high performance computing clusters where many of these simulations are run to capture all configuration and other files necessary to rerun simulations and reproduce their results.

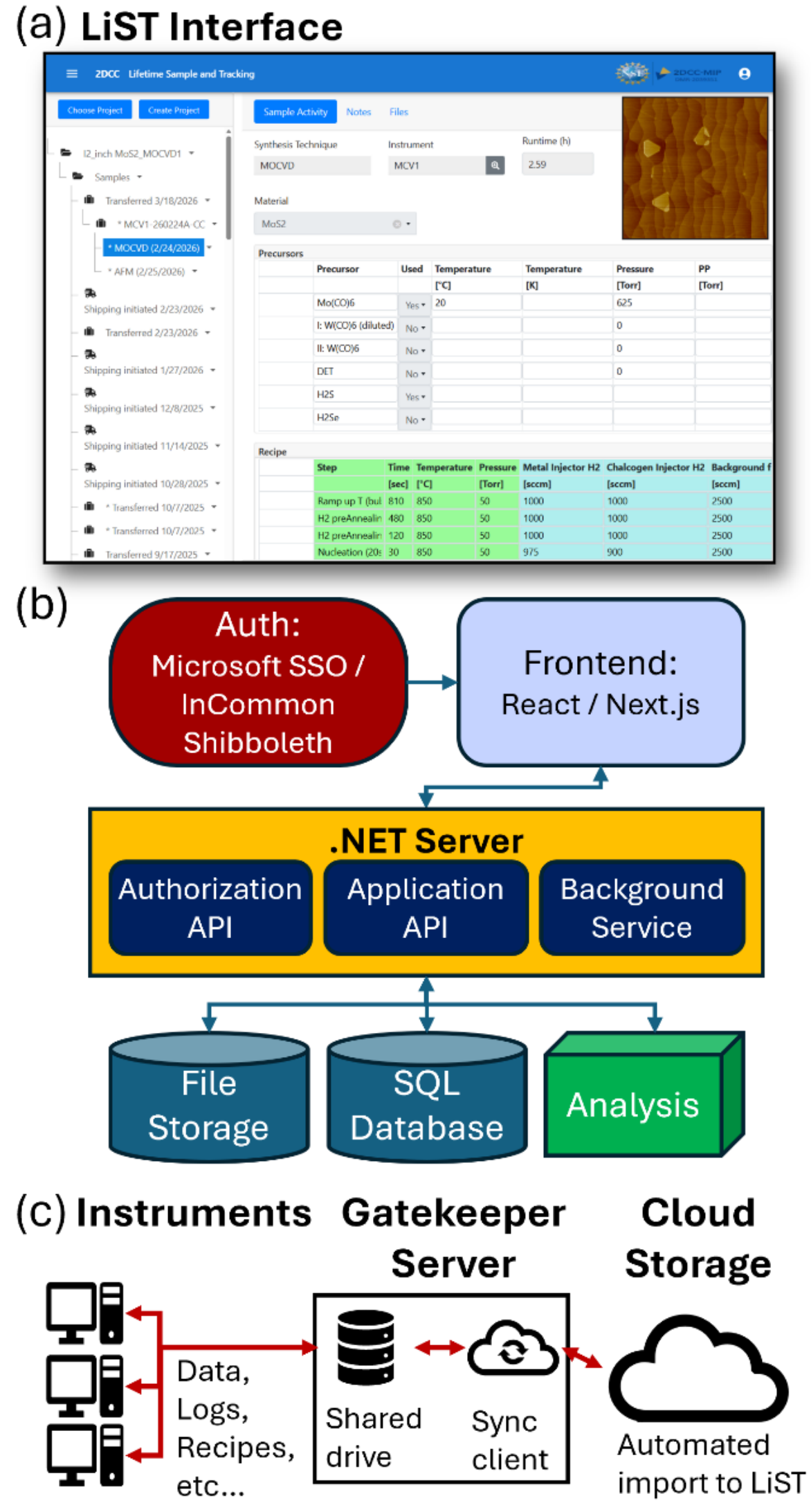

*Fig. 5 **(a)** The LiST web interface showing the growth parameters of an MoS2 sample and a corresponding AFM image in the inset. **(b)** Architecture of the LiST server, including authorization, web frontend, services and APIs. **(c)** Records from instruments are typically collected by writing to a shared drive on a gatekeeper server in each lab space that syncs them to cloud storage where they are ingested into LiST. This automation from data creation, to processing and archival is critical for capturing knowledge of the entire sample history.*

One reason an external 2DCC user might want to upload their own data into the LiST system is to utilize the Data Package functionality. These packages are created in LiST for inclusion in the data availability section of publications, containing all data used in the publication. This includes a list of all samples synthesized as part of the study reported in the publication, their growth recipes, and any additional characterization data that the authors wish to include. Prior to publication, a pre-publication link can be included in the manuscript allowing reviewers to view the raw data as part of the review process. On publication acceptance, the contents of the package are frozen, transferred to the Penn State ScholarSphere long-term data archive and made publicly available in the LiST application. The data package is then given a permanent DOI that resolves to the archive on ScholarSphere, as shown in Fig. 6(a). Use of ScholarSphere allows Penn State librarians to make sure the contents of the data packages are properly curated and documented. The ScholarSphere archive also includes a link back to LiST, where the data can be viewed within the LiST interface.

As the LiST system tracks which instruments were used to produce samples or generate characterization data, it utilizes another 2DCC developed tool called DMR-FIRST (Facilities and Instrumentation Research Search Tool) (https://dmr-first.org). While data packages aim to provide persistent DOIs to published datasets, DMR-FIRST provides persistent identifiers to scientific instrumentation. Similar to how ORCID persistent identifiers allow researcher publications to be tracked, DMR-FIRST DOIs allow instrument productivity and impact to be measured by including them in publications. In this way, LiST and DMR-FIRST are helping develop standards (Fig. 6(b)) that bridge missing links in documenting the research process. These could one day provide end-to-end provenance, assurance and attribution of data throughout the research lifecycle, as envisioned in the MGI goals of closing the loop in new materials development.

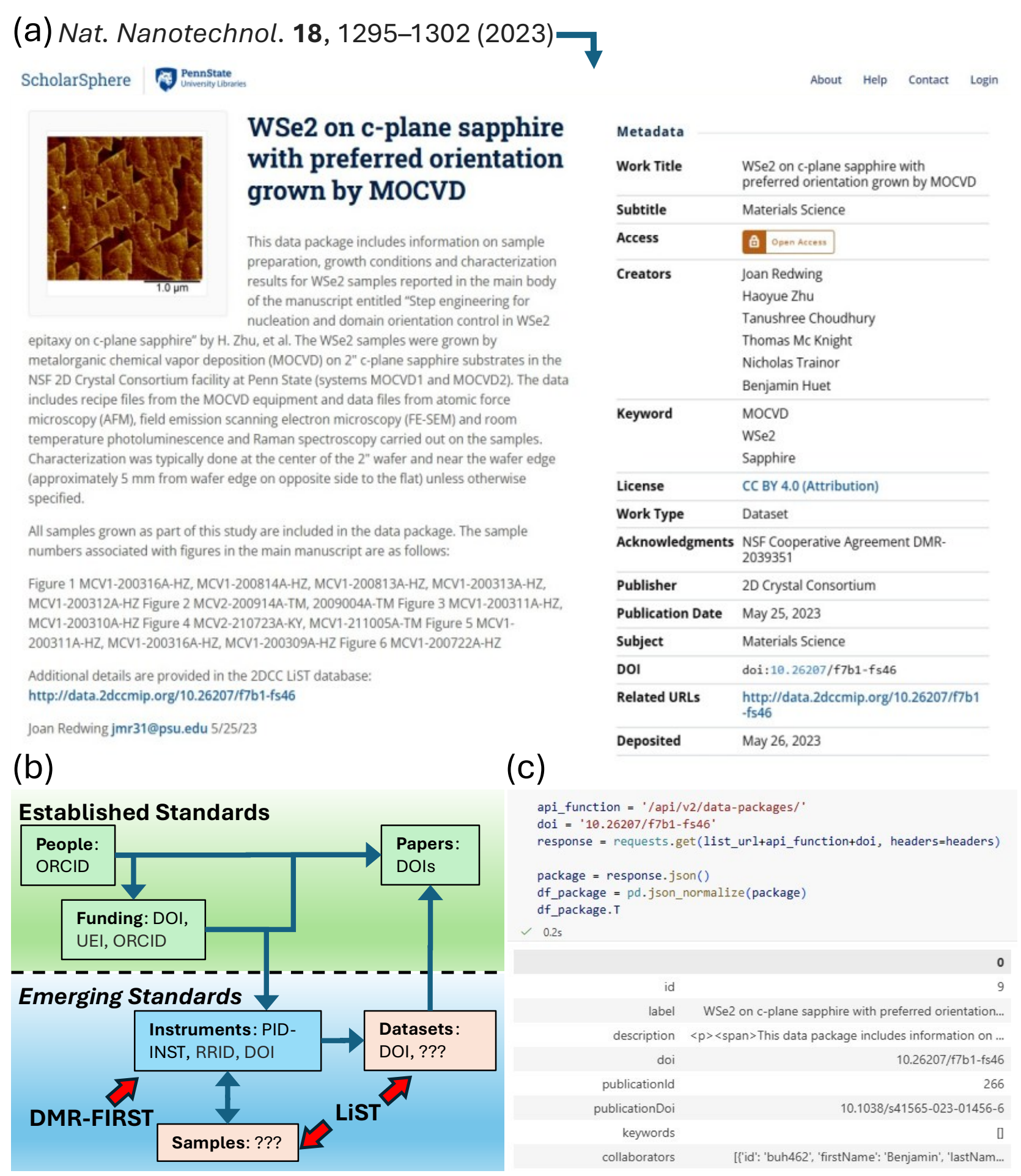

*Fig. 6 **(a)** Data packages are created in LiST to capture and preserve data used in publications such as Ref. 15. The data package contains growth parameters, characterization data and instrument identifiers associated with each sample used. It is archived in the Penn State University ScholarSphere long term repository and given a DOI that is linked to in the publication. ScholarSphere also has a Related URLs link where the data can be viewed in the LiST application. **(b)** The research process can be documented with linked entities to provide data providence and integrity. Standards for some basic entities such as samples and datasets are lacking, however. LiST Data Packages are datasets specifically relevant to publications and represent part of the 2DCC's commitment to make the research process more consistent with FAIR principles. **(c)** The LiST API can be used to query and retrieve Data Packages for reuse in machine learning or other applications.*

# Applications of LiST in Data-Driven Discoveries

## Feature detection using machine learning as a function of growth temperature

The study presented the application of transfer learning in a convolutional neural network (CNN) for classifying AFM images of $MoS_2$ grown at temperatures of 900 $^0$C, 950 $^0$C, and 1000 $^0$C (Fig. 7 (a))[14]. The dataset from LiST included approximately 260 topographs of $MoS_2$ grown using metal-organic chemical vapor deposition (MOCVD). The CNN utilized weights obtained from extensive training on the ImageNet dataset[16], which features everyday macroscopic objects, along with the MicroNet dataset[17], comprised of microscopy images. These pretrained weights were applied through feature extraction and fine-tuning of the model. Surprisingly, the results showed that using the pretrained weights from ImageNet outperformed those from MicroNet. This finding is counterintuitive, suggesting that the classification may depend more on image texture to determine the different growth temperatures of the materials. Model explanations were provided using class activation map (CAM) and occlusion attributions, both of which implicated crystal domains and qualities as temperature-dependent features.

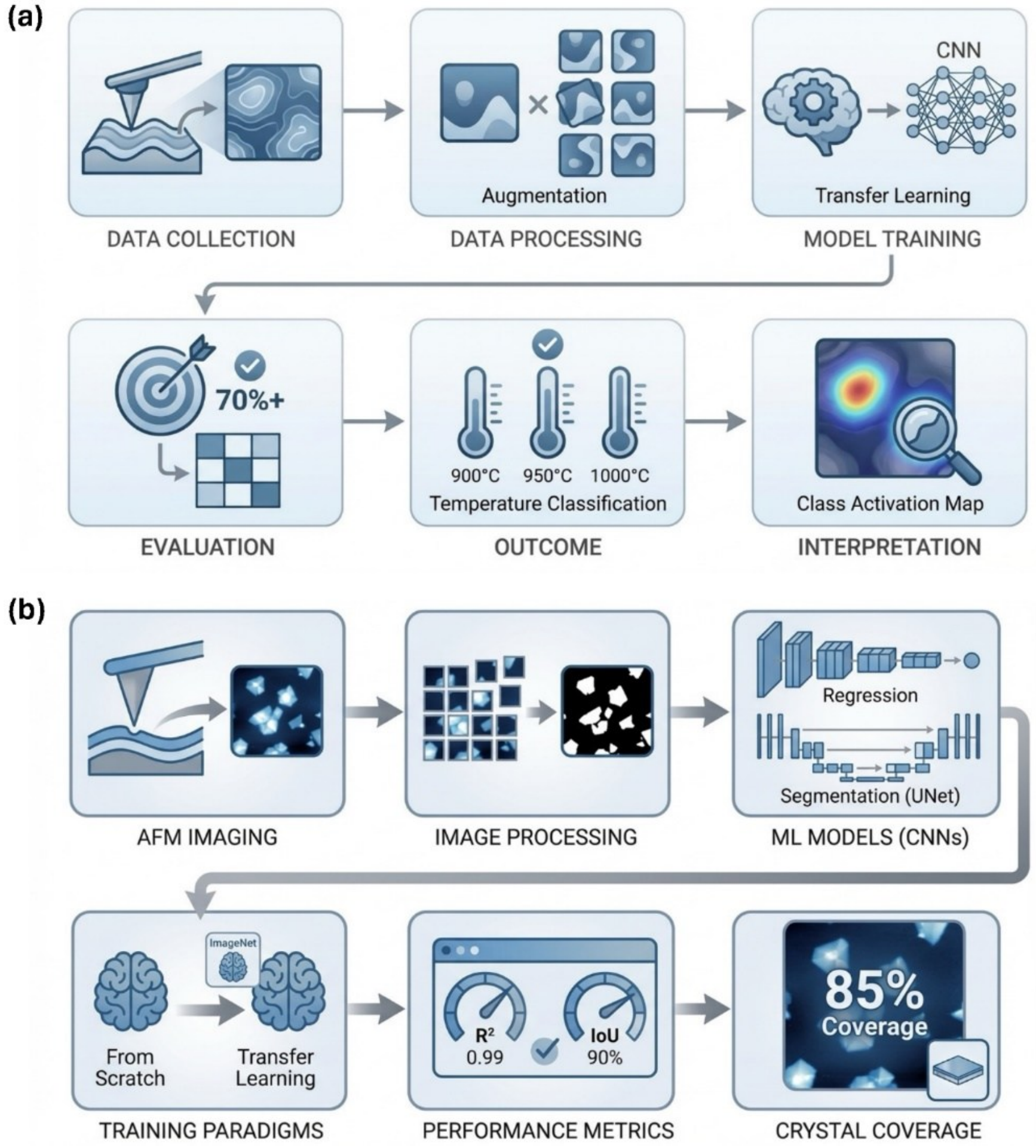

*Fig. 7 **(a)** illustrates transfer learning and machine learning algorithms for detecting features related to varying growth temperatures using AFM images of thin film MoS2 [14]. **(b)** presents regressions and segmentation models in determining the crystal coverage of MOCVD grown WSe2 thin film [11].*

### Determining crystal growth using segmentation and regression models

The study explored the application of semantic segmentation and regression models to assess the growth of crystals on substrates (Fig. 7(b))[11]. It utilized 52 micrographs of $WSe_2$ samples taken from the LiST. Transfer learning techniques from both ImageNet and MicroNet were employed for the regression and segmentation models, utilizing various CNN architectures. The findings indicated that ImageNet outperformed MicroNet in transfer learning when applied to the micrographs of transition metal dichalcogenides (TMDs). Additionally, the research demonstrated that regression models yielded better performance than segmentation models in measuring the percentage of the substrate surface covered by $WSe_2$ crystals.

**Transfer learning applied to multiple materials and varied data volumes of TMDs**

TMDs, including $MoS_2$, $WS_2$, $WSe_2$, $MoSe_2$, and Mo-$WSe_2$, were retrieved from the LiST to investigate various classification and transfer learning strategies applied to different volumes of data[10]. With approximately 1,000 samples of these five TMD materials, both multiclass and multilabel labeling schemes were explored. The results indicate that the multilabel approach provides a slight advantage in creating a generalizable model for material class predictions.

The transfer learning strategies examined included fine-tuning ImageNet weights to classify different material classes, whether two, three, four, or five classes. Additionally, a sequential transfer learning scheme was investigated (Fig. 8 (a)). In this approach, the fine-tuned ImageNet weights obtained from a smaller number of classes were further fine-tuned for classifying additional materials. For example, after fine-tuning the ImageNet weights to classify $MoS_2$ and $WS_2$ (two classes), those weights were subsequently fine-tuned to classify three materials: $MoS_2$, $WS_2$, and $WSe_2$. This sequential approach demonstrated better generalization to unseen samples, regardless of whether the dataset had low or high material volumes.

Furthermore, the trained models were interpreted, revealing that the latent features used to determine material classes correlated with measurable image morphological features, such as the difference of Gaussian (DoG) blob, root mean square (RMS) roughness, and grain density.

**Microscopy and spectroscopy cross-modal learning**

The study investigated how to project material properties from one type of measurement using data obtained from different tests[13]. It utilized AFM images, as well as Raman and photoluminescence (PL) spectra, of MOCVD-grown $MoS_2$. Approximately 170 data points, which included AFM, Raman, and PL measurements, were retrieved from the LiST database for analysis.

To uncover latent features, the study employed unsupervised representation of AFM images using ResNet pretrained models and Uniform Manifold Approximation and Projection (UMAP)[18]. Regression models were then developed to predict AFM features based on the Raman and PL spectra. Furthermore, generative models were utilized to produce complete Raman and PL spectra from the AFM data. The study also investigated how these spectra projected onto each other, as well as through the fusion of one spectrum with the AFM to generate the other spectrum (Fig. 8 (b)).

The findings indicated that the Raman spectra produced an effective predictive model for AFM features, whereas the PL spectra did not perform as well. Additionally, the AFM data showed strong results in generating both the Raman and PL spectra. Analysis of the results revealed that the mutual information between the data pairs increased in the following order: (PL, AFM), (Raman, AFM), and (Raman, PL).

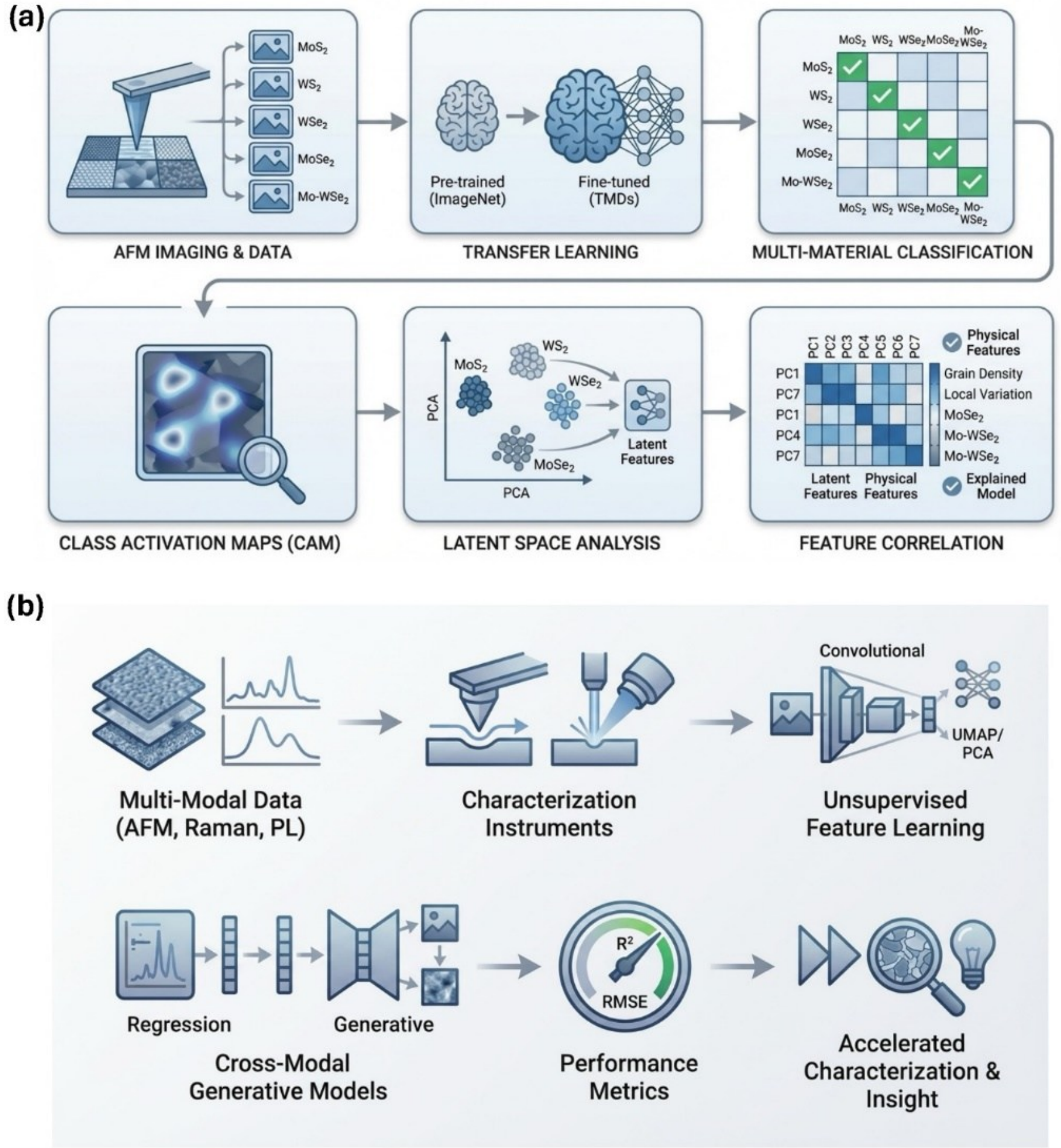

*Fig. 8 **(a)** presents different transfer learning and data labelling schemes for detecting features related to different classes of TMDs. Effect of data volume on model performance was also systematically studied[10]. Panel **(b)** presents multi- and cross-modal learning of AFM, Raman and PL of thin film MoS2 using unsupervised learning, regression, and generative autoencoder models[13].*

### Multimodal Machine Learning Analysis

A total of 37 thin-film GaSe samples were retrieved from the LiST for analysis, focusing on characterization data, RHEED, XRD, and AFM[19]. The study examined key growth parameters, including Ga flux, the Se:Ga flux ratio, and growth temperature. It also explored the low-dimensional representation of the RHEED data and the mutual information obtainable from the materials' quality metrics derived from this representation, as well as the other characterization data and growth parameters (Fig. 9). Moreover, the RHEED embedding and a regression model for the rocking curve full width at half maximum (FWHM) using the growth parameters were used to identify growth anomalies in the samples. Both approaches demonstrated good agreement in detecting anomalous samples.

### Machine Learning Model Interpretation using Synthetic Counterfactuals

The effect of growth parameters on material properties was studied using molecular beam epitaxy (MBE) grown SnSe[20]. We analyzed around 12 atomic force microscopy (AFM) samples of SnSe to investigate how the Sn:Se flux ratio and the order of precursor deposition—whether Sn was deposited first, Se first, or through co-deposition—affect the material's morphological properties. To examine how the flux ratio and the order of deposition affect various properties, we used a pretrained ResNet model along with Principal Component Analysis (PCA) to create low-dimensional representations of the AFM images. The PCA embedding indicated that the first and fifth components effectively represented the deposition order and flux ratio, respectively. We optimized the AFM images to identify features that could be adjusted through the PCA components. The generation of synthetic counterfactuals clarified the low-dimensional representation (Fig. 9), showing that isotropy and grain density were quantified by the first and second PCA components, respectively.

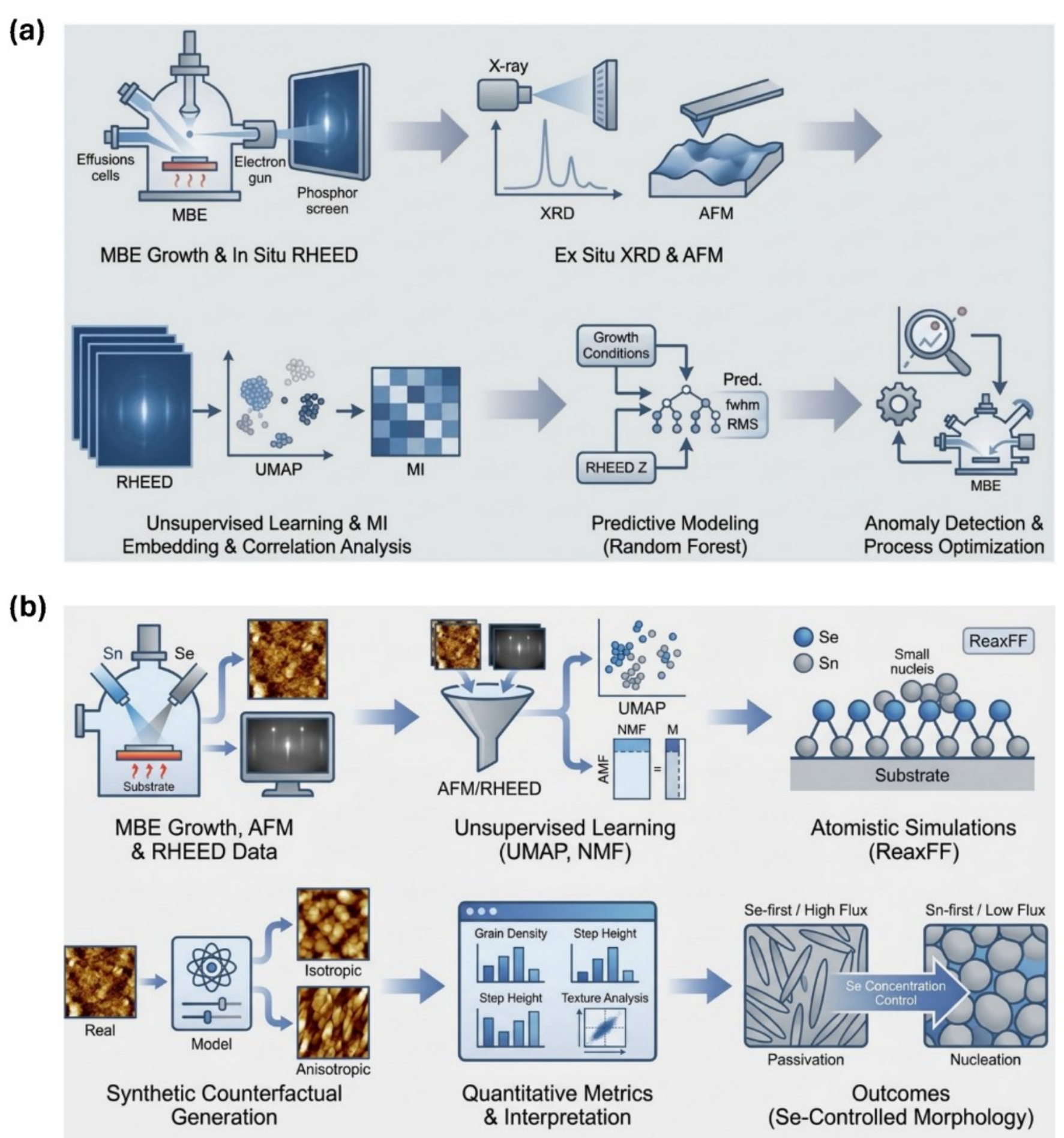

*Fig. 9* ***(a)*** *presents multi-modal analysis of the growth parameters and in-situ and ex-situ characterization data of GaSe synthesized with MBE. ML models deployed include unsupervised and regression models[19].* ***(b)*** *presents several approaches to understand and explain the effect of deposition order and the flux ratio in MBE synthesis of SnSe. Optimization-based counterfactual image generation were deployed to understand the effect of processes on the thin film morphology[20].*

# Conclusion

The 2DCC LiST data management platform can enhance materials research by helping researchers organize, preserve, and share their data. LiST is one of many tools being developed that can provide order in the often-chaotic landscape of materials science by fostering an ecosystem of open data and collaboration. This is highlighted by the LiST Data Package functionality that captures data and metadata of sample growth parameters, characterization data and analysis for materials used in publications, allowing greater transparency and enhancing replication in scientific literature.

Databases of experimental data that include both “good” and “bad” samples are crucial for training next generation models. Among these, we have demonstrated that latent code representations provide a powerful tool for understanding and quantifying spectral variations and their correlations with physical and morphological characteristics. Similarly, using pretrained models with AFM data can reveal insights into the structural and textural properties of $WS_2$ thin films that would only be possible with the wide and varied datasets that a system like LiST captures. The LiST system's connective characteristics between the front end, API, and backend database lay a foundation for the required tools needed to enable future research aspirations in autonomous synthesis via close coupling of *in-situ* characterization and synthesis instrumentation in one system.

The 2DCC is interested in opportunities for collaborations in sharing tools and developing metadata standards. Anyone that is interested in this area is encouraged to submit a Data Proposal at www.mri.psu.edu/2d-crystal-consortium.

# Acknowledgements

This study is based upon research conducted at The Pennsylvania State University Two-Dimensional Crystal Consortium – Materials Innovation Platform (2DCC-MIP) which was supported by NSF cooperative agreements DMR-1539916 and DMR-2039351, the center for 3D Ferroelectric Microelectronics and Manufacturing ($3DFeM^2$), an Energy Frontier Research Center funded by the U.S. Department of Energy, Office of Science, Office of Basic Energy Sciences Energy Frontier Research Centers program under Award Number DE-SC0021118 and the U.S. Department of Energy Office of Science National Quantum Information Science Research Centers as part of the Q-NEXT center that supported user interface enhancements and LiST database structure for MPCVD diamond growth.

# Author Declarations

**Conflict of Interest**

The authors have no conflicts to disclose.

**Author Contributions**

Konrad Hilse, Frank Santaguida, Ric Wilburn, Saiyyam Kochar developed the LiST application. Konrad Hilse and Anthony Richardella implemented the data syncing architecture. Anthony Richardella, Isaiah A Moses, Konrad Hilse and Kevin Dressler helped curate the data in the database. Kevin Dressler, Wesley F. Reinhart, Adri C. T. van Duin, Nitin Samarth, Vincent H. Crespi and Joan M. Redwing oversaw the technical direction of the project. Anthony Richardella, Konrad Hilse, Isaiah A Moses, Kevin Dressler, Wesley F.

Reinhart, Adri C. T. van Duin, Nitin Samarth, Vincent H. Crespi and Joan M. Redwing participated in the writing of this manuscript.

## Data Availability

Public datasets stored in LiST are available at: https://data.2dccmip.org

## References

1. Nayir, N. *et al.* Modeling and simulations for 2D materials: a ReaxFF perspective. *2d Mater.* **10**, 032002 (2023).

2. Wilkinson, M. D. *et al.* Comment: The FAIR Guiding Principles for scientific data management and stewardship. *Sci. Data* **3**, (2016).

3. Taylor, R. H. *et al.* A RESTful API for exchanging materials data in the AFLOWLIB.org consortium. *Comput. Mater. Sci.* **93**, 178–192 (2014).

4. Talirz, L. *et al.* Materials Cloud, a platform for open computational science. *Sci. Data* **7**, (2020).

5. Blaiszik, B. *et al.* The Materials Data Facility: Data Services to Advance Materials Science Research. *JOM* **68**, 2045–2052 (2016).

6. Zagorac, D., Muller, H., Ruehl, S., Zagorac, J. & Rehme, S. Recent developments in the Inorganic Crystal Structure Database: Theoretical crystal structure data and related features. *J. Appl. Crystallogr.* **52**, 918–925 (2019).

7. Chung, J. *et al.* SciStream: Architecture and Toolkit for Data Streaming between Federated Science Instruments. in *HPDC 2022 - Proceedings of the 31st International Symposium on High-Performance Parallel and Distributed Computing* 185–198 (Association for Computing Machinery, Inc, 2022). doi:10.1145/3502181.3531475.

8. Enders, B. *et al.* Cross-facility science with the Superfacility Project at LBNL. in *Proceedings of XLOOP 2020: 2nd Annual Workshop on Extreme-Scale Experiment-in-the-Loop Computing, Held in conjunction with SC 2020: The International Conference for High Performance Computing, Networking, Storage and Analysis* 1–7 (Institute of Electrical and Electronics Engineers Inc., 2020). doi:10.1109/XLOOP51963.2020.00006.

9. Stansberry, D., Somnath, S., Breet, J., Shutt, G. & Shankar, M. DataFed: Towards reproducible research via federated data management. in *Proceedings - 6th Annual Conference on Computational Science and Computational Intelligence, CSCI 2019* 1312–

1317 (Institute of Electrical and Electronics Engineers Inc., 2019). doi:10.1109/CSCI49370.2019.00245.

10. Moses, I. A. & Reinhart, W. F. Transfer learning for multi-material classification of transition metal dichalcogenides with atomic force microscopy. *Mach. Learn. Sci. Technol.* **5**, 045081 (2024).

11. Moses, I. A., Wu, C. & Reinhart, W. F. Crystal growth characterization of WSe2 thin film using machine learning. *Mater. Today Adv.* **22**, 100483 (2024).

12. Trice, R. *et al.* Machine Learning Guided Polymorph Selection in Molecular Beam Epitaxy of In2Se3. http://arxiv.org/abs/2601.13156 (2026).

13. Moses, I. A., Chen, C., Redwing, J. M. & Reinhart, W. F. Cross-Modal Characterization of Thin-Film $MoS_2$ Using Generative Models. *Advanced Intelligent Systems* **8**, 2500613 (2026).

14. Moses, I. A. & Reinhart, W. F. Quantitative analysis of MoS2 thin film micrographs with machine learning. *Mater. Charact.* **209**, 113701 (2024).

15. Zhu, H. *et al.* Step engineering for nucleation and domain orientation control in WSe2 epitaxy on c-plane sapphire. *Nat. Nanotechnol.* **18**, 1295–1302 (2023).

16. Deng, J. *et al.* ImageNet: A large-scale hierarchical image database. in *2009 IEEE Conference on Computer Vision and Pattern Recognition* 248–255 (IEEE, 2009). doi:10.1109/CVPR.2009.5206848.

17. Stuckner, J., Harder, B. & Smith, T. M. Microstructure segmentation with deep learning encoders pre-trained on a large microscopy dataset. *NPJ Comput. Mater.* **8**, 200 (2022).

18. McInnes, L., Healy, J. & Melville, J. UMAP: Uniform Manifold Approximation and Projection for Dimension Reduction. http://arxiv.org/abs/1802.03426 (2020).

19. Yu, M., Moses, I. A., Reinhart, W. F. & Law, S. Multimodal Machine Learning Analysis of GaSe Molecular Beam Epitaxy Growth Conditions. *ACS Appl. Mater. Interfaces* **17**, 34707–34716 (2025).

20. Chin, J. R. *et al.* Analyzing the impact of Se concentration during the molecular beam epitaxy deposition of 2D SnSe with atomistic-scale simulations and explainable machine learning. *Mater. Today Adv.* **28**, 100640 (2025).